\documentclass[a4paper,10pt]{article}
\pdfoutput=1
\usepackage{jcappub}
\usepackage{mathtools}
\usepackage{ulem}
\usepackage{bm}
\usepackage{xcolor}
\usepackage{subcaption}
\usepackage{float}
\usepackage{orcidlink}

\usepackage [latin1]{inputenc}

\renewcommand{\thesection}{\arabic{section}}

\def\cob{\color{blue}}

\newcommand{\au}[2]{#1.~#2}

\newcommand{\oarX}[1]{\href{http://arxiv.org/abs/#1}{{\ttfamily\cob arXiv:#1}}}
\newcommand{\arX}[1]{\href{http://arxiv.org/abs/#1}{{\ttfamily\cob arXiv:#1}}}
\newcommand{\doin}[6]{\href{http://dx.doi.org/#1}{{\cob {\it #2} {\bf #3 #4} (#6) #5}}}
\newcommand{\doinn}[5]{\href{http://dx.doi.org/#1}{{\cob {\it #2} {\bf #3} (#5) #4}}}
\newcommand{\doij}[5]{\href{http://dx.doi.org/#1}{{\cob {\it #2} {\bf #3} (#5) #4}}}

\newcommand{\tia}[1]{\textit{#1},}

\def\laq{~\raise 0.4ex\hbox{$<$}\kern -0.8em\lower 0.62ex\hbox{$\sim$}~}
\def\gaq{~\raise 0.4ex\hbox{$>$}\kern -0.7em\lower 0.62ex\hbox{$\sim$}~}

\def\beq{\begin{equation}}
\def\eeq{\end{equation}}
\def\bea{\begin{eqnarray}}
\def\eea{\end{eqnarray}}

\def \ra {\rightarrow}

\def \Mp {M_{\rm P}}

\def \da {\delta}
\def \b {\beta}
\def \a {\alpha}
\def \ap {\alpha^{\prime}}

\def \ga {\gamma}
\def \sg {\sigma}

\def \da {\delta}
\def \ep {\epsilon}
\def \r {\rho}
\def \om {\omega}
\def \Om {\Omega}

\title{Constraints on the  Pre-Big Bang scenario from a cosmological interpretation of the NANOGrav data}

\author[a]{P. Conzinu\,\orcidlink{0000-0002-7290-7790},}
\author[b]{G. Fanizza\,\orcidlink{0000-0001-5173-3800},}
\author[c]{M. Gasperini\,\orcidlink{0000-0001-9117-8303},}
\author[c]{E. Pavone\,\orcidlink{0000-0002-3022-4545},}
\author[c]{L. Tedesco\,\orcidlink{0000-0001-6508-2658}}
\author[d,e]{and G. Veneziano\,\orcidlink{0000-0003-3114-4894
}}

\affiliation[a]{Dipartimento di Scienze Matematiche, Fisiche e Informatiche, Universit\`a di Parma, \\
and INFN, Gruppo Collegato di Parma, Parco Area delle Scienze 7/A, I-43124, Parma, Italy}

\affiliation[b]{Dipartimento di Ingegneria, Universit\`a LUM, S.S. 100 km 18 - 70010
Casamassima (BA), Italy}
\affiliation[c]{
Dipartimento di Fisica, Universit\`a di Bari, 
Via G. Amendola 173, 70126 Bari, Italy,\\
and Istituto Nazionale di Fisica Nucleare, Sezione di Bari, Italy
}
\affiliation[d]{CERN, Theory  Department, CH-1211 Geneva 23, Switzerland}
\affiliation[e] {Coll\`ege de France, 11 Place M. Berthelot, 75005 Paris, France}

\emailAdd{pietro.conzinu@unipr.it}
\emailAdd{fanizza@lum.it}
\emailAdd{gasperini@ba.infn.it}
\emailAdd{eliseo.pavone@ba.infn.it}
\emailAdd{luigi.tedesco@ba.infn.it}
\emailAdd{gabriele.veneziano@cern.ch}

\abstract{ We discuss a recently proposed 
fit of the 15-year data set obtained from the North American Nanohertz Observatory for Gravitational Waves (NANOGrav) 
in terms of a relic stochastic background of primordial gravitons, produced in the context of the string cosmology pre-big bang scenario. We show that such interpretation cannot be reconciled with a  phenomenologically viable minimal version of such scenario, while it  can be allowed if one considers an equally viable but generalised, non-minimal version of pre-big bang evolution. Maintaining the $S$-duality symmetry throughout the high-curvature string phase is possible although somewhat disfavoured. The implications of this non-minimal scenario for the power spectrum of curvature perturbations are also briefly discussed.}

\keywords{Primordial gravitational waves (Theory), String cosmology, Pre-big bang
 
\vskip18pt 

\noindent{\bfseries\large\sffamily{Preprints:}}~CERN-TH-2024-210, BA-TH/809-24}

\begin{document}
\maketitle
\vskip 0.8 cm

\section{Introduction}  
\label{sec1}

There is an exciting possibility  that the signal observed by multiple Pulsar Timing Array (IPTA) collaborations, including NANOGrav \cite{2,2a}, the Parkes PTA (PPTA) \cite{2b,2c}, the European PTA (EPTA) in partnership with the Indian PTA (InPTA) \cite{2d,2e}, and the Chinese PTA (CPTA)\cite{2f}, can be interpreted as the first detection of a cosmological stochastic gravity-wave (GW) background. 

In particular, in a very recent paper \cite{1} the NANOGrav 15-year data set 
has been compared with the possible amplitude and frequency scale of the relic GW spectrum predicted long ago in the context of the pre-big bang (PBB) scenario \cite{3,4,Lidsey:1999mc, 5},   and thus used to constrain the related  string cosmology parameters\footnote{We recall that the PBB scenario is deeply rooted in the duality symmetries of the tree-level cosmological string equations \cite{6, Meissner:1991zj, Meissner:1991ge, Gasperini:1991ak, Sen:1991zi, 12}, which also constrain theoretically its parameters.}. 

The results presented in \cite{1} are certainly interesting, but they seem to disagree with another recent analysis of the primordial spectrum of relic pre-big bang gravitons \cite{7}. 
In fact, let us recall that  the data fit presented in \cite{1} is based on a string cosmology GW spectrum $\Om_{\textsc{gw}}(f)$ with two different frequency branches. In the low-frequency brach, \(\Omega_{\textsc{gw}}\) scales as \(f^3\) up to a transition frequency \(f_s\), characteristic of string theory and marking the onset of the high-curvature regime. Beyond \(f_s\), the spectrum retains a power-law form but with a different exponent: specifically, \(\Omega_{\textsc{gw}} \sim f^\alpha\), where \(\alpha < 3\).

The data fit  presented in \cite{1} indicates that, for frequencies $f_s \laq f  \laq 10^{-6}$ Hz,  
the high frequency spectrum is nearly flat or slightly decreasing   
and that the spectral amplitude satisfies 
\beq
\Om_{\textsc{gw}} (f_{s}) \simeq 2.9^{+5.4}_{-2.3} \times 10^{-8}, ~~~~~~~~~~~~~~~~~
f_{s} \simeq (1.2\pm 0.6) \times 10^{-8} {\rm Hz}.
\label{11}
\eeq
Such a spectral amplitude is {\it outside} the allowed region of the plane $\{ f, \Om_{\textsc{gw}}\}$, determined in \cite{7} on the grounds of a phenomenologically complete model of pre-big bang evolution.

The main purpose of this paper is twofold. First of all we will explain in Sect. \ref{sec2} why the two discussions of the pre-big bang spectrum presented in \cite{1} and \cite{7} lead to different results, even if both are correct within their own assumptions. Second, we will suggest in Sect. \ref{sec3} how the minimal scenario used in \cite{7} could be generalised  to accommodate the production of a relic signal  that is also consistent with the data fit  presented in (\ref{11}).
Finally, Sect. \ref{sec4} will be devoted to some concluding remarks. 

\section{A phenomenologically viable minimal Pre-Big Bang scenario} 
\label{sec2}

The first point to be stressed is that the relic GW spectrum discussed in \cite{1} refers to a preliminary and very simple example of pre-big bang spectrum presented in various old papers \cite{8,9,10,11}. Such a spectrum, however, may be regarded as incomplete as it does not include all frequency branches arising in a realistic and phenomenologically viable string model of the early Universe.

A viable inflationary scenario must indeed predict, besides the relic GW background, also a related spectrum of adiabatic scalar curvature perturbations able to explain the observed large scale anisotropy. This has been shown to be possible \cite{12,13,14} (see also \cite{15} for a more complete review) thanks to the contribution of the string Kalb-Ramond axion field acting as a curvaton \cite{16,17}, and producing a cosmic phase of axion dominated oscillations. The inclusion of this important aspect in the scenario modifies however the standard post-bouncing evolution and this, in its turn, may affect the sub-horizon propagation of tensor perturbation modes after their re-entry, thus producing additional frequency branches with different spectral indices in today's observed GW spectrum (see e.g. \cite{7,18,19} for recent quantitative discussions of such an effect). 

The pre-big bang model used in \cite{1} does not take into account this crucial ingredient determining the final observed GW spectrum, and thus neglects the important associated constraints. Such an ingredient, on the contrary,  is properly accounted for in the spectrum analyzed in \cite{7}. As a result, the corresponding, more tightly constrained spectral region does not overlap with the region of the IPTA signal (see, in particular, Fig. 4 of \cite{7}).
It should be stressed, however, that the spectral model considered in \cite{7} is based on a complete but ``minimal" example of pre-big bang scenario. It corresponds to the simplest description of a complete bouncing evolution from the string perturbative vacuum down to the present epoch, in agreement with all string theory constraints (see e.g. \cite{Conzinu:2023fth} for an exact  string model of bounce), but is also based on ad hoc assumptions chosen to minimise the number of unknown parameters. It may be useful to recall here, in view of its generalisation to be presented in the next section, the basic aspects of such a minimal scenario. 

First of all it must include, for its completeness, at least two different pre-bouncing phases as well as other two post-bouncing phases, occurring before the reheating epoch marking the beginning of standard cosmology.  Starting from initial conditions asymptotically approaching the flat perturbative vacuum with vanishing Hubble parameter, $ H \ra 0$, such a scenario is thus characterized by four different Hubble scales: $H_s$, marking the beginning of the string high-curvature regime; $H_1$, corresponding to the bouncing from accelerated to decelerated, decreasing curvature expansion; $H_\sg$, marking the beginning of the dust-like phase dominated by the oscillating axion; and $H_d$, associated to the axion decay  that triggers the reheating and the beginning of the standard radiation-dominated era. Obviously,
\beq
H_s \laq H_1 \laq \Mp, ~~~~~~~~~~~~~~H_1 \gaq H_\sg > H_d \gaq H_N,
\label{21}
\eeq
where $H_N \simeq (1 {\rm MeV})^2/\Mp$ is the scale of standard nucleosynthesis, and $\Mp \equiv (8 \pi G)^{-\frac12} \simeq 2\times 10^{18}$ GeV is the (reduced) Planck mass scale. We also recall, for later use, that $H_\sg$ and $H_d$ can be expressed in terms of the axion mass $m$ and of the initial, post-bouncing axion amplitude $\sg_i$ as follows  \cite{13,14,15}:
\beq
H_\sg \simeq m \left(\sg_i\over \Mp\right)^4, ~~~~~~~~~~~
H_d \simeq m \left(m\over \Mp\right)^2.
\label{22}
\eeq
The axion field starts oscillating at a scale $H\simeq m$, hence the condition that the axion is oscillating when it becomes dominant (required for the curvaton mechanism to be efficient) implies $m \gaq H_\sg$, namely\footnote{This is  effectively the case if, as expected, the axion potential is periodic with a periodicity smaller than $\Mp$.}  
$\sg_i \laq \Mp$.   

The amplification of metric perturbations, in this scenario, is thus characterised by four typical frequency scales: $f_s, f_1, f_\sg, f_d$, where $f_s$ is the proper frequency of a mode crossing the horizon at the beginning of the string phase, $f_1$ is the maximal amplified frequency, while $f_\sg$ and $f_d$ are the frequency of modes re-entering the horizon, respectively, at  the beginning and  conclusion of the axion-dominated era. It follows, automatically, that $f_1 \gaq f_\sg > f_d$, while $f_s$ satisfies $f_s \laq f_1$, but its particular value is free, in principle, with respect to the values of $f_\sg$ and $f_d$. We shall assume here, as in \cite{7,18,19,Conzinu:2023fui}, that $f_s$ is smaller than the other frequencies typical of the pre-big bang scenario, but larger than the frequencies constrained by the CMB data, i.e. the {pivot} frequency scale $f_* \simeq 0.05\, {\rm Mpc}^{-1}$, and the typical frequency of Large Scale Structure (LSS) observations, $f_{\rm LSS} \simeq 60\, f_*$. 
Hence, the model we shall consider here will be characterised by the following hierarchy of frequency scales:
\beq
f_* < f_{\rm LSS} \laq f_s \laq f_d < f_{\sg} \laq f_1.
\label{23}
\eeq

Given an inflationary scenario with four typical frequency scales, the amplified spectrum of Fourier modes of tensor perturbations, $h_k$, will be characterised in general by four different spectral branches to be computed by solving, for each mode $k$, the canonical perturbation equation
\beq
v_k'' +\left(k^2 - {\xi_h''\over \xi_h} \right) v_k=0.
\label{24}
\eeq
Here $v_k= \xi_h(\eta) h_k$ is the (Mukhanov-Sasaki) variable \cite{19a} for which the effective action for the tensor field $h_k$ takes the standard canonical form, the (background dependent) variable $\xi_h(\eta)$ is the so-called pump field controlling the GW dynamics in the various cosmic phases, and a prime denotes differentiation with respect to the conformal time $\eta$. It should be recalled that the Fourier parameter $k$ for a mode re-entering the horizon at a given time $t$ in the post-bouncing epochs described by a standard FLRW metric background, is related to the  proper frequency of that mode observed at the present time $t_0$, i.e. $f(t_0)$, by $f(t_0) = \om(t_0)/2 \pi$, where $\om(t_0)= k/a(t_0)= H(t) a(t)/a(t_0)$.

In order to solve Eq. (\ref{24}) and obtain the GW spectrum, we need the explicit behaviour of the pump field for tensor perturbations in the various cosmic phases. The answer is simple for the low-curvature (low energy) regimes, where we can use the tree-level string cosmology equations. Assuming, as in \cite{7,18,19,Conzinu:2023fui}, to start with a ten-dimensional gravi-dilaton string background with 3 expanding dimensions with scale factor $a$, and 6 contracting spatial dimensions with scale factors $b_i$, which asymptotically evolves from the perturbative vacuum at $\eta \ra - \infty$ up to the string scale (i.e. for $0 \leq H\leq H_s$), we then find for the tensor pump field the simple form
\beq
\xi_h(\eta) \sim  a  g_4^{-1} = a \left( \prod_{i=1}^6 b_i \right)^{1/2} e^{-\phi/2},
\label{24a}
\eeq
where 
$\phi$ is the dilaton and $g_4$ is the effective (time-dependent) 4-dimensional string coupling. It is well known \cite{29a,29b} that this 
 uniquely fixes the power law behaviour of $\xi$ to be $\xi_h(\eta) \sim  (-\eta)^{1/2}$. 
 
The answer is simple also in the post-bouncing regime if we assume that in our minimal scenario, at all epochs from $H_1$ down to the present epoch $H_0$, the internal dimensions as well as the dilaton, controlling the string coupling, are already stabilised at the string scale. In that case the tensor pump field simply coincides with the scale factor $a(\eta)$ describing decelerated, decreasing curvature expansion. In the radiation-dominated phases, occurring from $H_1$ to $H_\sg$ and for $H_d \geq H \geq H_{\rm eq}$, where $H_{\rm eq}$ is the equality scale, the pump field is then given by  $\xi_h(\eta) \sim\eta$, while for the dust-like phases, i.e. for $H_\sg > H > H_d$ and  $H<H_{\rm eq}$ we have, as usual, $\xi_h(\eta) \sim \eta^2$.

Finally, for the pump field of the high-curvature string phase, i.e. for $H$ ranging from $H_s$ to $H_1$, we can still use a parametrisation based on a power-law behaviour, but we have to take into account the effects of higher order string $\ap$ corrections, as well as other possible high energy effects typical of string theory. In the minimal scenario considered in \cite{7,18,19,Conzinu:2023fui} it has been assumed $\xi_h(\eta) \sim (-\eta)^{-1+\b}$, where the factor $(-\eta)^{-1}$ corresponds to having frozen the string-frame
curvature  at  the string scale\footnote{Attractors of this type have been discussed in \cite{20}.} and, $\b$ is a positive parameter describing the rate of growth of the four-dimensional string coupling $g_4$ (a combined effect of the dilaton and  internal volume time-dependence, see Eq. (\ref{24a})) according to:
\beq
\b  \equiv  \frac {d \log g_4}{d \log a} ~;~ ~~~~~~~~~~~~~~ 0 \laq \b <3.
\label{25}
\eeq
Here the lower limit is required for a monotonically growing coupling, while the upper limit is to avoid quantum background instabilities \cite{21}. The idea of the scenario is that $g_4$ starts very small at the beginning of the string phase and becomes of ${\cal O}(1)$ at its end.

It is important to stress that the same parameter $\b$, in this scenario, also appears (with the opposite sign) in the axion pump field $\xi_\sg$ governing the amplification of the axion fluctuation during the string phase, which takes in fact the form  $\xi_\sg(\eta) \sim  (-\eta)^{-1-\b}$, according to the $S$-duality symmetry of string theory \cite{12}. Hence, the same parameter $\b$ appears in the primordial spectrum $P_s(f)$ of scalar curvature perturbations produced via the curvaton mechanism, and thus contributes to  the important constraint following from the standard normalisation of the scalar spectrum at the CMB pivot scale (see e.g. \cite{32a}), which implies \cite{7,18,19,Conzinu:2023fui}
\beq
P_s(f_*) \equiv 2.1 \times 10^{-9} \simeq {T^2 (\sg_i)\over 2 \pi^2} \left(H_1\over \Mp\right)^2 \left(f_s\over f_1\right)^{3-|3+2\b|} \left(f_*\over f_s\right)^{n_s-1},
\label{26}
\eeq
and which provides a stringent constraint on all the parameters. Here $n_s \simeq 0.965$ is the scalar spectral index, and 
$T(\sg_i)$ is the transfer function connecting the amplitude of axion and scalar curvature perturbations, which can be written (according to a numerical integration of the perturbation equations \cite{14}) as
\beq
T(\sg_i)\simeq 0.13 \left(\sg_i\over \Mp \right)+ 0.25 \left(\Mp\over \sg_i\right) -0.01.
\label{27}
\eeq

Summing up and imposing all above mentioned constraints (given by Eqs. (\ref{25}), (\ref{26}),  plus the hierarchy of scales (\ref{23}), plus the condition $\sg_i \leq \Mp$), it turns out that the allowed amplitude of the relic GW spectrum for this minimal scenario (see \cite{7}) cannot reproduce
the results (\ref{11}) obtained with the fit of the NANOGrav data set (unless we allow $\beta <0$, which is however inconsistent with the physical interpretation of this parameter, see Eq. (\ref{25})). We have checked that the same result is obtained even if we assume higher (and in principle allowed) values of the frequency $f_s$, changing the hierarchy of Eq. (\ref{23}) and choosing, for instance,  $f_d \laq f_s <f_\sg$, or $f_\sg \laq f_s \laq f_1$.


\section{A non-minimal  Pre-Big Bang scenario} 
\label{sec3}

\subsection{Parametrization of the non-minimal model and related constraints}
\label{non minimal model}	

The minimal scenario of the previous section takes into account two typical effects of the high-curvature string phase: the late-time attractor and the growth of the dilaton. But there are in principle other possible high-energy string theory effects like, for instance, the production of a dense  gas of primordial, string-size, black holes or ``string holes" \cite{Veneziano:2003sz, 22,Bitnaya:2023vda}. Such effects can modify not only the background evolution but also, and in a different way, the propagation of different types of perturbations like 
tensor-metric \cite{Dolgov} and axion-field perturbations.

It seems appropriate, therefore, to consider also a more general, non-minimal phenomenological scenario where, during the high-curvature string phase, the tensor and axion pump fields ($\xi_h$ and $\xi_\sg$) can still be described by a power-law behaviour but with two new parameters, in principle unrelated to $\beta$ and also to each other (in case $S$-duality is broken). We can parametrize them  as follows:
\beq
\xi_h \sim  \left(-\eta\right)^{-1+\b +\ga} \equiv  \left(-\eta\right)^{-1+\beta_h};~~~~
\xi_\sg \sim \left(-\eta\right)^{-1-\b +\da} \equiv  \left(-\eta\right)^{-1+\b_\sg};~~~\b_\sg = - \b_h + \ep, 
~~~~\epsilon \equiv \da + \ga \, .
\label{31}
\eeq
We have $\epsilon =0$ if the S-duality assumed in the minimal scenario is still valid, while we recover the previous scenario if both $\gamma$ and $\da$ are vanishing. Let us then present the modified spectrum, and the related constraints, for the non-minimal case with $\ga \not= 0$ and $\da \not= 0$. 
{In this paper we will discuss the possible range of values of $\da$ and $\ga$ allowed by present phenomenological constraints. However, let us stress that their precise values, and in particular the value of the parameter $\ep$ controlling the possible breaking of the S-duality symmetry and the related time behaviour of the effective string coupling, should be computed, and physically interpreted, on the grounds of a given explicit model of background evolution (as we a replanning to discuss in a future paper).}

Assuming the same hierarchy of frequency scales as before  (given by Eq. (\ref{23})), solving Eq. (\ref{24}) in the various phases (with the new tensor pump field), matching the solutions, and computing the final, presently observed GW spectral energy density expressed in units of critical density $\rho_c$, i.e. $\Om_{\rm GW}(k,t_0) = \r_c^{-1}(t_0) d\r_k(t_0)/d \ln k$, we obtain (besides a negligible contribution for $f > f_1$):

\beq
	\!\!\!\!\!\!\!\!\!\!\!\!\!\!\! \!\!\!\!\!
	\frac{\Om_{\textsc{gw}}(f,t_0)}{\Om_{\textsc{gw}}(f_1,t_0)}  = 
	\begin{cases}
        \left(\dfrac{f}{f_1}\right)^{3- |3-2 \b_h|}, &f_\sg \laq f \laq f_1 \\ \\
		\left(\dfrac{f_\sg}{f_1}\right)^{3- |3-2 \b_h |}
		\left(\dfrac{f}{f_\sg}\right)^{1- |3-2 \b_h|},\qquad  &f_d \laq f \laq f_\sg \\ \\
		 \left(\dfrac{f_\sg}{f_1}\right)^{3- |3-2 \b_h |} \left(\dfrac{f_d}{f_\sg}\right)^{1- |3-2 \b_h |}
		\left(\dfrac{f}{f_d}\right)^{3- |3-2 \b_h |},\qquad
        &f_s \laq f \laq f_d\\ \\
 \left(\dfrac{f_\sg}{f_1}\right)^{3- |3-2 \b_h |}\left(\dfrac{f_d}{f_\sg}\right)^{1- |3-2 \b_h |}
\left(\dfrac{f_s}{f_d}\right)^{3- |3-2 \b_h |}
		\left(\dfrac{f}{f_s}\right)^{3},\qquad \,\,\,& f \laq f_s\,\,.
	\end{cases}
	\label{32}
\eeq
 Note the difference  from the results of \cite{7} due to  the new parameter $\ga$ (or rather $\b_h$). 

In order to discuss the various phenomenological constraints it is useful  to work, as in \cite{7}, with the following frequency ratios
\beq
z_s ={f_1\over f_s}, ~~~~~~~
z_d ={f_1\over f_d}, ~~~~~~~
z_\sg ={f_1\over f_\sg}, ~~~~~~~~~~~~~
z_s \gaq z_d > z_\sg \gaq 1.
\label{33}
\eeq
In terms of such variables, the end-point amplitude of the spectrum is given by \cite{7}:
\beq
\Om_{\textsc{gw}}(f_1,t_0)= \Om_r(t_0) \left(H_1\over \Mp\right)^2 \left(z_\sg\over z_d\right)^2,
\label{34}
\eeq
where $ \Om_r(t_0)\sim 10^{-4}$ is the present critical fraction of radiation energy density (including neutrinos), and we have neglected a possible suppression of $\Om_{\rm{GW}}(f_1,t_0)$ due to significant late entropy production \cite{10}. Also, the typical axion parameters $\sg_i$ and $m$, controlling the post-bouncing scales $H_\sg$ and $H_d$ according to Eq. (\ref{22}), can be written as:
\beq
{m\over \Mp} \simeq \left(H_1\over \Mp\right)^{1/3} z_d^{-1} z_\sg^{1/3},
~~~~~~~~~~~~~~~~~~
{\sg_i\over \Mp} \simeq \left(H_1\over \Mp\right)^{1/6} z_d^{1/4} z_\sg^{-7/12}.
\label{35}
\eeq

The constraints to be imposed on this scenario can now be explicitly written (in base $10$ logarithmic form, useful for later applications) as follows. 
The condition $\sg_i \laq \Mp$ becomes:
\beq
\log \left(\frac{H_1}{\Mp}\right) + {3\over 2} \log z_d \laq {7\over 2} \log z_\sg.
\label{36}
\eeq
The condition $H_d \gaq H_N$ becomes: 
\beq
\log \left(\frac{H_1}{\Mp}\right)- 3 \log z_d + \log z_\sg \gaq \log\left(\frac{H_N}{\Mp}\right)\approx-42 - \log 4\,.
\label{37}
\eeq
The condition $f_{\rm LSS} < f_s$  (see Appendix B.2 of \cite{7}) becomes:
\beq
\log z_s \laq 26 - \log 9 +\frac{1}{2} \log\left(\frac{H_1}{\Mp}\right)+\frac{1}{2} \left( \log z_\sg - \log z_d \right)\,.
\label{38}
\eeq
The phenomenological normalisation (\ref{26}), generalized to the non-minimal scenario, leads to a condition which also include the new parameter $\da$ (or $\ep$, see Eq. (\ref{31})): 
\bea
\log \left(\frac{H_1}{\Mp}\right) &=& \frac{2}{5 - n_{\rm s}} 
\left\{ \log \left[\frac{4.2 \pi^2}{T^2(\sg_i)}\right] -9 + (1-n_{\rm s})(\log  1.5 - 27)\right.\nonumber \\
&&  + \left. \left(4-n_{\rm s} - |3+2(\beta_h-\epsilon) | \right) \log z_s +\frac{n_{\rm s}-1}{2} \left(\log  z_\sg - \log z_d \right) \right\} \laq 0,
\label{39}
\eea
where the inequality on the right hand side has been imposed according to Eq. (\ref{21}). 

All the constraints listed up to now are  to be always applied, in general,  for the internal as well as for the phenomenological consistency of the non-minimal scenario that we are considering. But let us now introduce a further, more constraining ingredient, by imposing on the lowest energy branch of our non-minimal spectrum (\ref{32}) to exactly satisfy the numerical values of Eq. (\ref{11}), to be in agreement with the data fit of \cite{1}.
From the first condition of Eq. (\ref{11})  we then obtain
\beq
2 \log \left(\frac{H_1}{\Mp}\right)\simeq -4 +\log 2.9 +\left(3 - |3-2\b_h|\right)
\log z_s, 
\label{310}
\eeq
while the second condition of Eq. (\ref{11}) gives 
\beq
{1\over 2}\log \left(\frac{H_1}{\Mp}\right)\simeq -19 + \log 1.93 + 
\log z_s
 +{1\over 2} \left(\log z_d - \log z_\sg \right),
\label{311}
\eeq
that makes the inequality of Eq. \eqref{38} automatically satisfied. In such a context one should note that the amplitude $\Om_{\rm{GW}}(f_s)$ 
of Eq. (\ref{11}), reached by the spectrum at a relatively low frequency scale $f_s \sim 10^{-8}$ Hz, is quite close to the experimental upper bound recently imposed by the data of the LIGO-Virgo-KAGRA (LVK) network \cite{31}, which provides the condition:
\beq
\Om_{\textsc{gw}}(f_{\rm LVK})< 4.12 \times 10^{-8} ,~~~~~~~~~~~~
f_{\rm LVK} \simeq 35.4\,{\rm Hz}\,. 
\label{312}
\eeq
Taking into account that our spectrum (\ref{32}) may be flat, or even increasing, for $f>f_s$, it follows that we should add to the list of our constraints also the above condition, suitably imposed on the whole range of spectral frequencies larger than $f_s$.

In addition, since the string phase should be characterized by a curvature scale of order the string mass scale $M_{\rm S}$, we should in principle restrict the ratio $\left({H_1}/{\Mp}\right) $ to lie in the canonical range $10^{-2} - 10^{-1}$. Nonetheless, in order to consider also some more exotic possibilities, and/or to consistently take into account the synthetic definition of  our bouncing parameter $H_1$ (where we have absorbed model dependent numerical factors), 
we will enlarge our parameter space to include the wider range:
\beq
-3 \laq \log \left(\frac{H_1}{\Mp}\right) \laq -1. 
 \label{314}
 \eeq
{We stress that such a slight extension of the possible range of the bouncing scale is by no means compulsory to obtain models compatible with the data fit of Eq. (\ref{11}) (as will be shown in the following subsections). Also, it does not necessarily imply deviations from the natural value of the string mass scale if the bouncing is also controlled by the growth of the string coupling and by a (possibly non local) dilaton potential. Finally, such an extension is simply suggested by the phenomenological approach adopted in this paper, and we are leaving a more  constrained and predictive discussion to the case of explicit models that will be presented in forthcoming papers.}

We should also take into account the possibility that, in the non-minimal scenario we are considering, the modified axion parameter $\b_\sg$ is small enough but positive, thus describing a spectrum of induced scalar-curvature perturbations which is growing at high frequency. In that case the lowest frequency branch $f \laq f_s$ of the scalar spectrum has then the same behaviour reported in Eq. (\ref{26}) (with $\b$ obviously replaced by $- \b_\sg$), while for $f \gaq f_s $ we find \cite{13,14}
\beq
P_s(f) \simeq {T^2 (\sg_i)\over 2 \pi^2} \left(H_1\over \Mp\right)^2 \left(f\over f_1\right)^{3-|3-2\b_\sg|}  ,
~~~~~~~~~ f_s \laq f \laq f_1.
\label{314a}
\eeq
In such a case we have to impose a further constraint, needed for the self-consistency of our scenario: the condition of negligible backreaction of the produced perturbations on the assumed model of background evolution, $P_s(f)\laq 1$,which, imposed at the peak values of the scalar spectrum, gives the condition:
\beq
\log {T (\sg_i)\over\sqrt{2 \pi^2}}+ \log \left(\frac{H_1}{\Mp}\right) \laq 0. 
\label{314b}
\eeq
By using our previous result $\sg_i \laq \Mp$ we can also rewrite the above condition in simplified form as $H_1 \laq \sg_i$. 

The question is now: is it possible to find a set of parameters $\{\b_h, \b_{\sg}, z_s, z_d, z_\sg , \sg_i, m, H_1\}$  satisfying all physical and model-dependent constraints  of the non-minimal scenario, Eqs. (\ref{25}), (\ref{33}), (\ref{36}), (\ref{37}), (\ref{38}), (\ref{39}), (\ref{312}), and compatible with the two  conditions (\ref{310}), (\ref{311}) obtained from the analysis of the NANOGrav data, plus the additional conditions (\ref{314}),  (\ref{314b})?

\subsection{Allowed region in parameter space}
\label{region}	

The total number of parameters of our non-minimal (yet simple) model is quite large. The whole set consists of $ H_1, m,\sg_i, z_s, z_d, z_\sg$ and the two spectral parameters $\b_h$ and $\b_{\sg}$.  However, they are not all independent. After several attempts, we have found the best way to present our results to be as follows.

We first choose a value for $\sg_i$. Recalling that, before a non-perturbative axion potential is generated, any value of $\sg_i$ (modulo its periodicity ${\cal O}( \Mp)$) is equally probable, we could sample, for instance, the values ${\sg_i}/{\Mp}=1, 0.5, 0.1$. Although much smaller values of $\sg_i$ may correspond to considerable fine-tuning of initial conditions, we will allow for the more generous range:
\beq
10^{-3/2} \laq \frac{\sg_i}{\Mp} \laq 1. 
 \label{sgi}
 \eeq
In any case, as we shall see, an order-of-magnitude change in $\sg_i$ has only a modest effect on the allowed ranges for the other parameters. 
{We recall indeed that $\sg_i$, the initial value of axion background after the bounce, controls the details of the curvaton mechanism (as discussed in \cite{13,14}), and defines different initial configurations depending of course on the explicit model of bounce, 
but, in general, it is not affected by the dynamics of the phase of pre-big bang evolution.}

\begin{figure}[H]
    \centering
    \begin{subfigure}[t]{0.45\textwidth}
        \centering
        \includegraphics[width=\textwidth]{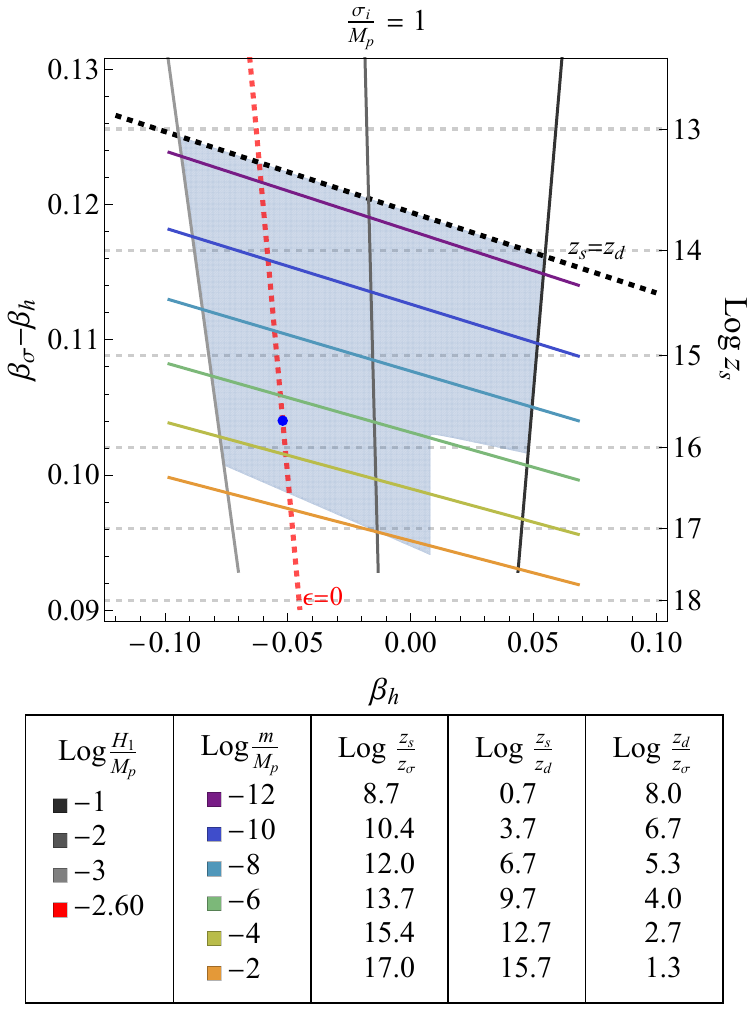}
        \caption{}
        \label{fig:plot1}
    \end{subfigure}
    \hfill
    \begin{subfigure}[t]{0.45\textwidth}
        \centering
        \includegraphics[width=\textwidth]{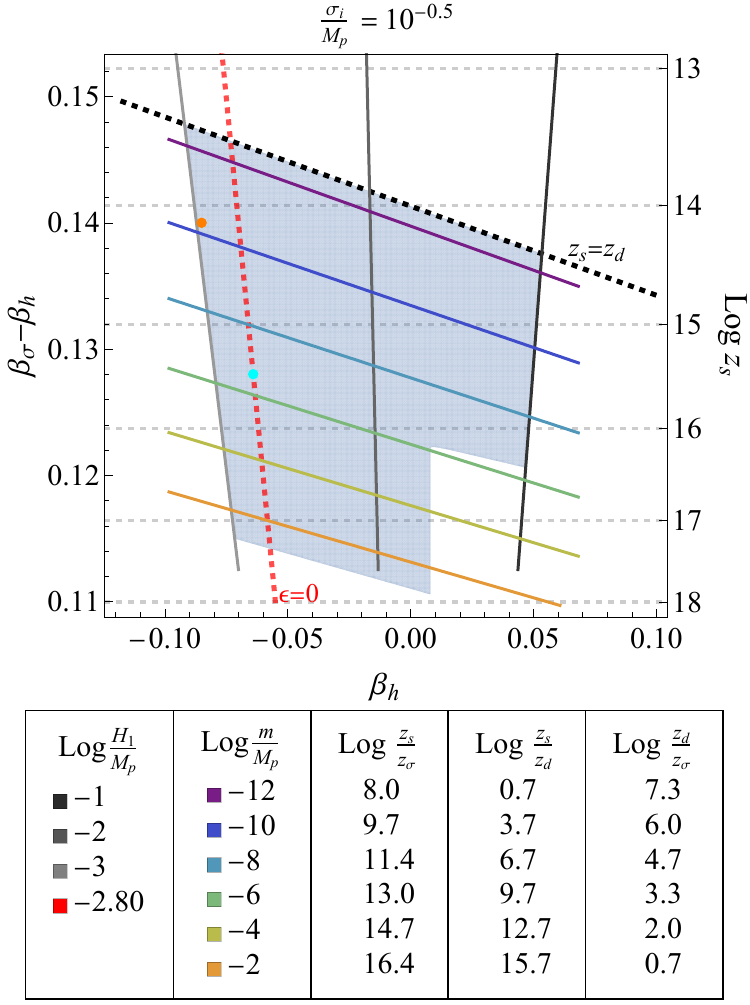}
        \caption{}
        \label{fig:plot2}
    \end{subfigure}
      \hfill
    \begin{subfigure}[t]{0.45\textwidth}
        \centering
        \includegraphics[width=\textwidth]{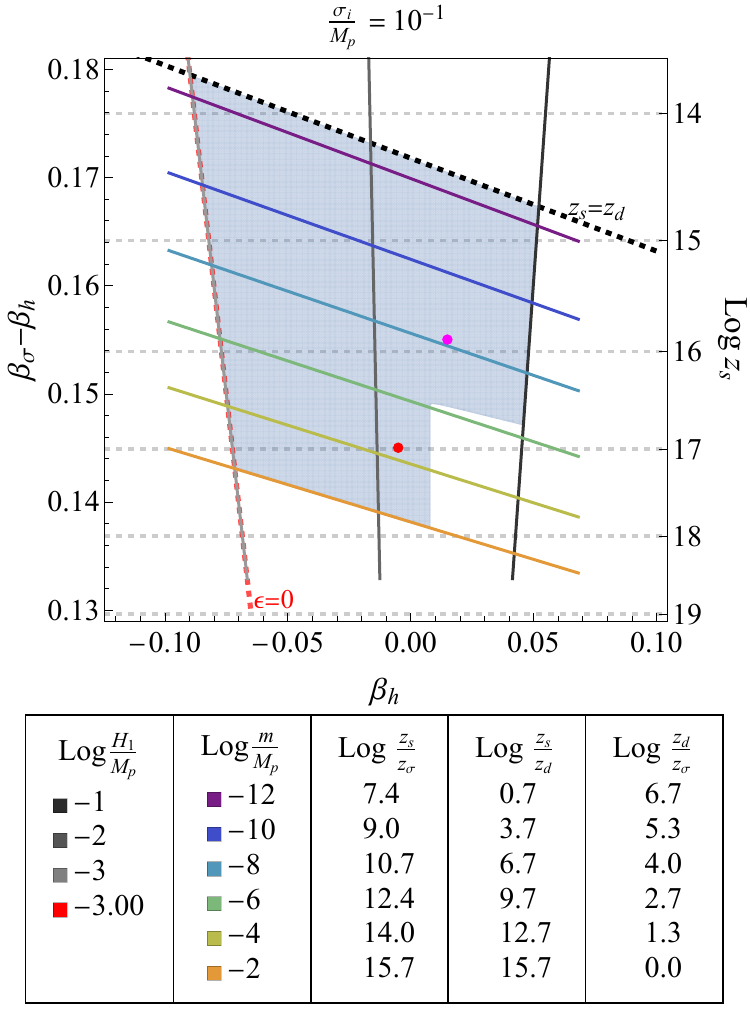}
        \caption{}
        \label{fig:plot2}
    \end{subfigure}
    \hfill
    \begin{subfigure}[t]{0.45\textwidth}
        \centering
        \includegraphics[width=\textwidth]{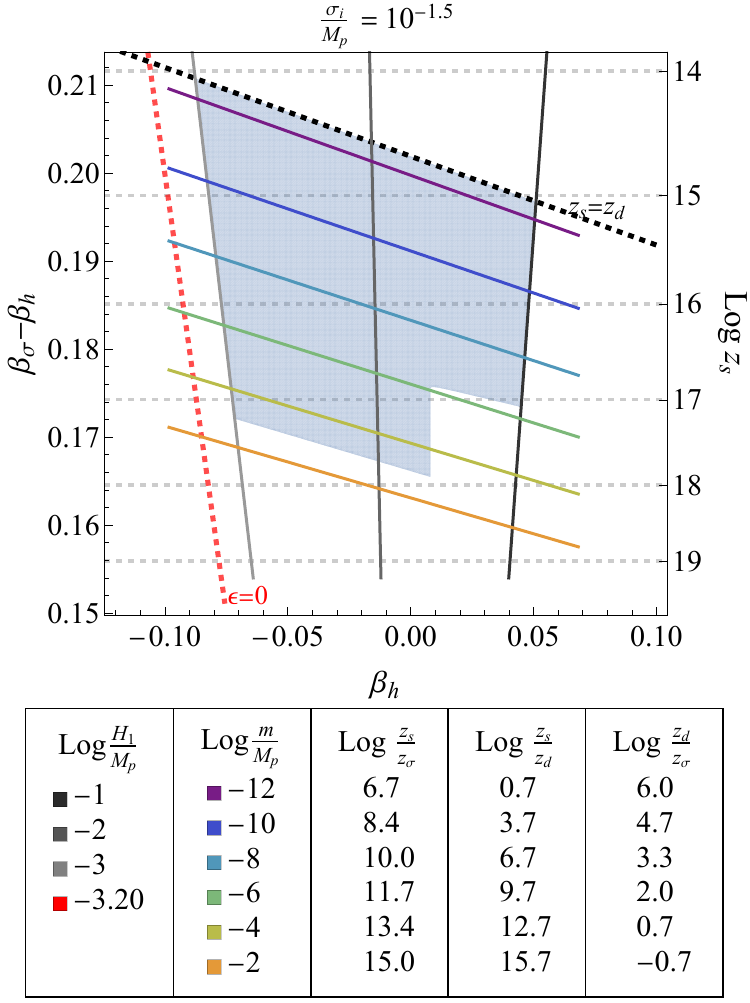}
        \caption{}
        \label{fig:plot2}
    \end{subfigure}
    \caption{The shaded blue region represents the allowed parameter space. The condition $\epsilon = 0$ (red dashed line) defines the duality region, where $\beta_h = -\beta_\sigma$. Vertical black lines correspond to variations in the parameter $\log\left(H_1 / M_p\right)$, while horizontal dashed gray lines represent the parameter $\log z_S$. The dashed black lines, marking the curve $z_s =z_d$, delineate the upper boundary of the allowed region. Oblique curves illustrate variations in $\log\left(m / M_p\right)$, $\log\left(z_s / z_d\right)$, $\log\left(z_s / z_\sigma\right)$, and $\log\left(z_d / z_\sigma\right)$, as explained in the legend. The colored dots correspond to the GW spectra reported in Fig.~\ref{f2a}.
    }
\label{fig:Parameter_Space}
\end{figure}

Once $\sg_i$ is given, there are enough equations (i.e. (\ref{35}), (\ref{39}), (\ref{310}), (\ref{311})) to determine all the remaining physical parameters in terms of the two  spectral ones,  $\b_h$ and $\b_{\sg}$. Therefore, the  constraints (\ref{314}),(\ref{sgi}), (\ref{36}), (\ref{37}),(\ref{38}) define a (hopefully non empty) region in an easily plotted $\{\b_h, \b_{\sg}\}$ plane, that will be illustrated in Fig. \ref{fig:Parameter_Space}.

The two missing constraints are the ones given by LVK bound (\ref{312}) and by the absence of strong back-reaction effects (\ref{314b}). It is easy to check that this latter constraint is automatically satisfied in the region already defined (indeed, for ${H_1}/{\Mp} \laq 0.1$, it only requires ${\sg_i}/{\Mp} > 0.01$).  
On the other hand, the LVK bound can be  non trivial (depending on the sign of $\b_h $),  and cuts off some corner of parameter space as we shall discuss in a moment. Within the resulting allowed region, each point represents a consistent spectrum of gravitational and scalar perturbations that can be readily drawn as we shall see in the next subsection.
Inside the allowed region we can draw contour lines along which each of the remaining parameters $(H_1, m, z_s, z_d, z_\sg)$ takes constant values (so-called level curves).  Luckily, one finds that  $z_s$ (the total duration of the string phase in red-shift space) is only a function of the combination $\b_{\sg}- \b_h$. This explains why it is convenient to plot our parameter space in the two-dimensional plane spanned by the coordinates $\left\{\b_h, \b_{\sg}- \b_h\right\}$, as we have done in Fig. \ref{fig:Parameter_Space}.

Two further simplifications occur when solving our equations, at least for  $3-2\b_\sg > 0$ (which turns out to be largely satisfied by our constraints). These are apparent in the explicit solutions given in Eqs. (\ref{rel1}) in the Appendix~\ref{secA}.
The first is that the level-curves of $H_1/M_p$ are straight lines coming from the origin $\b_{\sg} = \b_h=0$. The second is that the level curves of ${m}/{\Mp},\,{z_s}/{z_d},\,{z_s}/{z_{\sg}}$ are all the same (they are given by the straight lines originating from the point $\b_{\sg} = \b_h=2$). 
That also means that, for a given ${\sg_i}/{\Mp}$ and ${m}/{\Mp}$,  the two ratios ${z_s}/{z_d}$ and ${z_s}/{z_{sg}}$ can be determined. 
For the convenience of the reader we also give in the Appendix \ref{secA} these explicit relations.

After the above discussion it is now straightforward to describe the properties the allowed region of parameter space  shown in the four panels of Fig.~\ref{fig:Parameter_Space}, 
corresponding to the four choices  $\log({\sg_i}/{\Mp}) = 0, -1/2, -1, -3/2$. We stress immediately that the region has a trapezoidal shape except for a small ``tooth" on the lower right side. This is precisely  the extra constraint due to the LVK bound on $\Omega_{\textsc{gw}}$, physically due to the fact that the value of $\Omega_{\textsc{gw}}(f_s)$ imposed to fit the NANOGrav data is close to the upper limit of LVK; thus, for a positive $\beta_h$, it is not trivial to avoid a clash between the two constraints. 

In each panel the $\sg_i$-dependent  correspondence between $\b_{\sg} - \b_h$ and $z_s$ is clearly displaced. The level curves for $\log({H_1}/{\Mp})=-1,-2,-3$ are the nearly vertical straight lines. To these we add the line (dotted in red) 
representing the condition $\ep=0$, i.e. $\b_\sg = -\b_h$, and corresponding to models satisfying the $S$-duality symmetry. In the duality-symmetric case the corresponding value of $\log ({H_1}/{\Mp})$ is a mildly varying function of $\sg_i$ and is indicated in the accompanying Table. It varies between $-2.60$ and $-3.20$ for our chosen interval of $\sg_i$ and is within the allowed region for ${\sg_i}/{\Mp} > 0.1$, a value with 90\% probability of being realized 
assuming uniform priors for the parameters distribution.

Finally, the nearly horizontal lines are the common level curves for the remaining  quantities ${m}/{\Mp}, {z_s}/{z_d}, {z_s}/{z_\sg}$ whose corresponding values are also given again in each Table (together with their trivial combination ${z_d}/{z_\sg}$).
It may be useful to note, also, that relaxing the lower limit (\ref{314}) of $H_1$ has the effect of allowing higher and higher values of $\b_\sg$, as well as lower and lower (negative) values of $\b_h$. The same effect on the parameter $\b_\sg$ is obtained if we relax the lower limit (\ref{sgi}) on the axion amplitude $\sg_i$; in that case, however, the related effects on $\b_h$ are much smaller and, in any case, the allowed range of $\b_h$ does not increase, but it tends to sligthly decrease. 

Such properties are important to understand the variation in shape of the GW spectrum under a given variation of its parameters, to be illustrated in the next Sect.~\ref{spectra} where we will present some examples of GW spectra coming from the various regions of Fig.~\ref{fig:Parameter_Space}. 
In the final subsection~\ref{sec31} we will also present, for completeness, the associated power spectra of induced scalar curvature perturbations, leaving to future works the study of their possible implications .

\subsection{Typical spectra for non-minimal models consistent with a fit of the NANOGrav data}
\label{spectra}	

\begin{figure}[t]
\centering
\includegraphics[width=14cm]{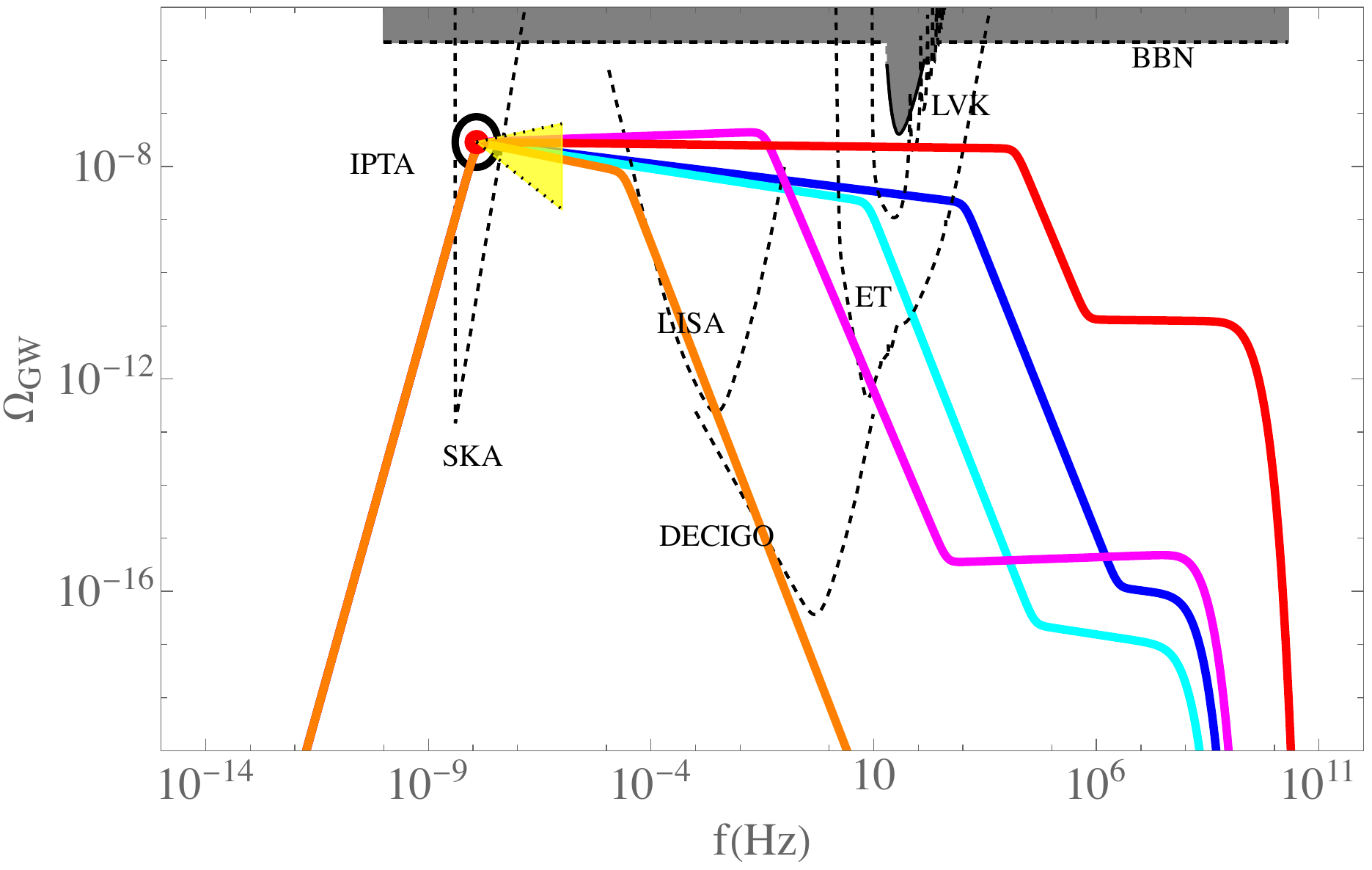}
\caption{
Possible examples of relic primordial GW spectrum produced  in the context of a non-minimal model of pre-big bang evolution satisfying all present phenomenological constraints, and consistent with the data fit reported in Eq. (\ref{11}) 
(the red dot localised inside the black circle). The corresponding values of their parameters are shown in Table~\ref{tab:parameters}. Also shown  are the expected sensitivity of SKA, LISA, ET, DECIGO (the regions inside the dashed curves), and the upper bounds (the grey shaded areas) imposed by the present results of the LVK network and by the standard nucleosynthesis. 
The blue and cyan curves  are example of spectra compatible with $S$-duality symmetry, while the red, magenta and orange curves are  obtained if such a symmetry is violated in the high-curvature string phase. The yellow shaded region describes the allowed spectral region for $f_s < f < 10^{-6}$ Hz suggested by the results of \cite{1}. }
\label{f2a}
\end{figure}

Given the allowed values of the parameters defined in the previous subsection, we can now easily illustrate the possible spectral distribution of the relic GW background (\ref{32}) in the various frequency branches. 
In spite of the rather small and compact size of the allowed region of parameter space there is a wide range of possible spectra that, we have illustrated in Fig. \ref{f2a} in order to compare their shape with the expected sensitivity of present and near future GW detectors. We have plotted the spectra using a smooth interpolation between the various branches\footnote{The smoothing of the piecewise profile (\ref{32}) does not change the underlying scenario because
the transition epochs from one phase to another are of negligible duration  compared with the time extension of such phases.}, following the method previously introduced in \cite{7} 
(in particular, in Appendix A).
{The different sets of parameters producing the various spectra of this section have been chosen to emphasise the possible differences in the spectral index and in the frequency extension of the intermediate frequency branch, $f_s \leq f \leq f_d$. As we shall see, such differences have important phenomenological implications.}

For our illustrative purpose we have plotted indeed in the $\{f,\Om_{\textsc{gw}}\}$ plane a few spectra which are all compatible with the bounds of Fig.~\ref{fig:Parameter_Space} but  which describe distinct physical configurations,  such as different extensions of a large spectral amplitude towards the high frequency range (the red and orange curves), or the possibility of a growing behaviour for modes with $f > f_s$, amplified during the high-curvature string phase (the magenta curve).

We have also included, for comparison, examples with a similar spectral behaviour but produced by models preserving the $S$-duality symmetry, and thus characterised by a parameter $\ep=0$ (the blue and cyan curves). It may be interesting to note that, for all duality invariant models, the high-frequency spectral branches with $f > f_s$ must be characterised by a decreasing behaviour, given the negative allowed range of $\b_h <0$ along the red dotted lines $\ep=0$ (see Fig.~\ref{fig:Parameter_Space}). The precise numerical values of all the parameters for the five plotted spectra are listed in Table~\ref{tab:parameters}.

In any case, it should be noted that all spectra  have their lowest-frequency branch well inside the expected sensitivity of the Square Kilometer Array (SKA) \cite{ska}  collaboration and that, for most of the spectra,  the peak value turns out to be localised just in correspondence of the value reported in Eq. (\ref{11}) and obtained from International Pulsar Timing Array (IPTA) collaboration (IPTA) \cite{24a}, as illustrated in the figure. Note also that the high-frequency behaviour of the spectrum may by compatible (in agreement with the results of \cite{7}) 
with the expected sensitivity range of near-future detectors, represented by the regions inside the dashed curves of Fig.~\ref{f2a}: in particular those of LISA \cite{27}, ET \cite{28}, DECIGO \cite{29} and SKA \cite{ska}.
Marginally, also with the expected sensitivity of Advanced LIGO \cite{30}.
In addition, all the plotted spectra are automatically compatible with the well known bound on the standard big bang nucleosinthesys (BBN) \cite{10}, which requires $\Om_{\textsc{gw}} < 2.2 \times 10^{-6}$ in a very wide frequency range $f \gaq 10^{-8}$ Hz, and which is illustrated by the shaded grey area of Fig.~\ref{f2a}. 
\begin{table}[h!]
\centering
\begin{tabular}{|c|c|c|c|c|c|c|c|c|}
\hline
\textbf{} & $\log z_s$ & $\log z_d$ & $\log z_\sigma$ & $\epsilon$ & $\beta_h$ & $\log \frac{\sigma_i}{M_{\text{P}}}$ & $\log \frac{H_1}{M_{\text{P}}}$ & $\log \frac{m}{M_{\text{P}}}$ \\ \hline
\textcolor{magenta}{\rule{10mm}{2pt}} & $15.89$ & $9.44$  & $5.32$ & $0.185$  & $0.015$  & $-1.00$   & $-1.53$ & $-8.18$ \\ \hline
\textcolor{red}{\rule{10mm}{2pt}}     & $16.98$ & $4.88$  & $3.28$ & $0.135$  & $-0.005$  & $-1.00$  & $-1.85$ & $-4.41$ \\ \hline
\textcolor{orange}{\rule{10mm}{2pt}}  & $14.14$ & $10.84$ & $4.65$ & $-0.03$ & $-0.085$ & $-0.50$   & $-2.97$ & $-10.30$ \\ \hline
\textcolor{blue}{\rule{10mm}{2pt}}    & $15.70$ & $4.74$  & $1.29$ & $0$     & $-0.052$ & $0.00$   & $-2.60$ & $-5.17$ \\ \hline
\textcolor{cyan}{\rule{10mm}{2pt}}    & $15.46$ & $6.68$ & $2.93$ & $0$     & $-0.064$ & $-0.50$  & $-2.80$ & $-6.62$ \\ \hline
\end{tabular}
\caption{Numerical values of the parameters for the spectra plotted in Fig.~\ref{f2a}}
\label{tab:parameters}
\end{table}

Finally, it may be interesting to check that the allowed spectra of our non-minimal model are compatible not only with the normalisation (\ref{11}) of the spectral amplitude, but also with the power-law behaviour $f^\a$ suggested at the $90 \%$ confidence level by the data fit of \cite{1}  in the frequency range $f_s \laq f \laq 10^{-6}$ Hz, with a power roughly given by $-0.66 \laq \a \laq 0.18$.  The corresponding allowed region for the spectrum  is illustrated by the yellow shaded area of Fig. \ref{f2a}, well consistent with the allowed spectra of the non-minimal  models both with and without $S$-duality symmetry.

\subsection{Remarks on the spectrum of scalar curvature perturbations} 
\label{sec31}	

To complete our presentation of the main physical aspects of the non-minimal scenario, introduced in order to support a possible cosmological interpretation of the IPTA  signal, it may be useful to briefly illustrate also the properties of the primordial spectrum of adiabatic scalar curvature perturbations produced by the axion through the curvaton mechanism \cite{13,14,16,17}, and associated to the relic GW spectrum discussed before.    

Let us recall, to this purpose, that in the minimal pre-big bang scenario the (superhorizon) scalar spectrum of metric perturbations $P_s(f)$ at the axion decay time $\eta_d$ is simply proportional to the primordial axion perturbation spectrum $P_\sg(f)$ at all perturbation scales, i.e. $P_s \sim T^2 P_\sg$, where $T$ is given by Eq. (\ref{27}). Also, in the minimal scenario, the low frequency branch of the scalar spectrum ($f<f_s$) has a slightly decreasing power-law behaviour in agreement with the observed CMB anisotropy, i.e. $P_s(f) \sim f^{n_s-1}$. For the high frequency modes, leaving the horizon during the high energy string phase ($f_s < f < f_1$), the power-law behaviour is determined by the same parameter $\b$ (but with the opposite sign) controlling the growth of tensor perturbations, so that $P_s(f) \sim f^{-2\b}$. Since $\b>0$ it turns out, for the minimal scenario, that the intensity of the scalar spectrum is always decreasing with frequency, and it becomes fully negligible in the high frequency limit. 

In the context of the non-minimal scenario that we are considering here the situation is quite similar, but with only one (crucial) difference: the two parameters controlling the pump field evolution and the spectral distribution, in the case of tensor perturbations ($\b_h$) and of axion/scalar perturbations ($\b_\sg$), are in general different and in principle unrelated, $\b_\sg = - \b_h + \ep$, see Eq. (\ref{31}). However, they are strongly constrained by the whole set of theoretical as well as phenomenological conditions discussed in Sects. 3.1, 3.2. It turns out, in particular, that the allowed values of $\b_\sg$ are always positive, and small enough so that the high frequency branch of the scalar perturbation spectrum, according to Eq. (\ref{314}), is always growing in frequency as $P_s(f) \sim f^{2\b_\sg}$, $\b_\sg >0$.

Note that, unlike the GW spectrum, the axion and (as a consequence) the curvature power spectrum do not have breaks in the slope at $f_d$ and $f_\sg$. This is because axion perturbations re-entering the horizon during the axion-dominated phase are already non relativistic (since $ f/(2\pi) \sim  H < m$) as discussed in detail in Appendix A of  \cite{14}.

\begin{figure}[t]
\centering
\includegraphics[width=12cm]{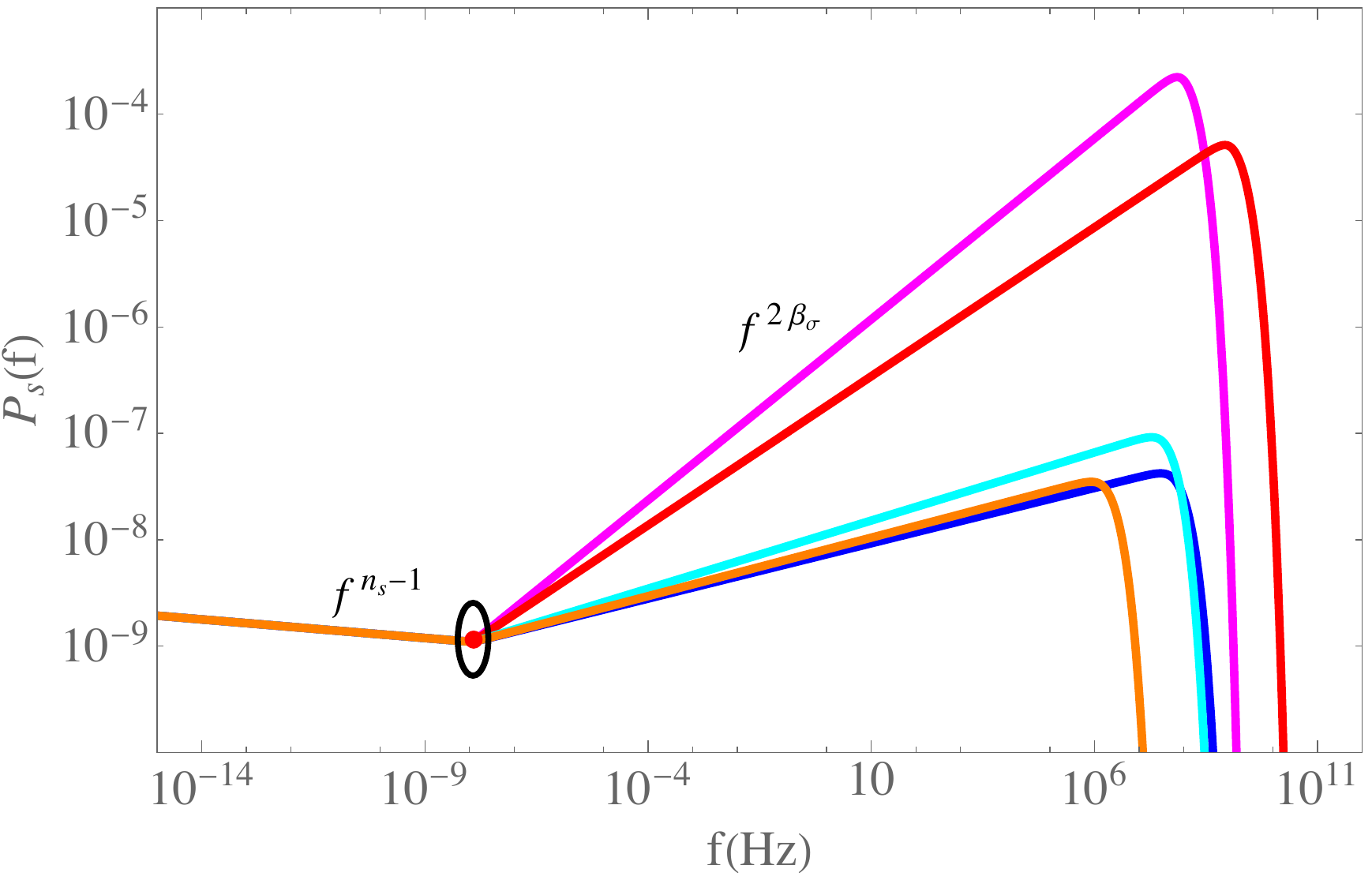}
\caption{The five spectra of scalar perturbations associated with the five GW spectra of the non-minimal models illustrated in Fig.~\ref{f2a} and Table \ref{tab:parameters} (with the same colors).The change of slope correspond to the frequency $f_s$ obtained from the fit of the NANOGrav data. Note that, contrary to the case of the GW spectrum, the scalar spectrum is always growing for $f>f_s$.}
\label{f3}
\end{figure}

This important physical difference between the minimal and non-minimal scenario is emphasized in Fig.~\ref{f3}. We have plotted, for the non-minimal scenario, the primordial  spectra of scalar perturbations exactly corresponding to the GW spectra  illustrated in Fig.~\ref{f2a}, and with the colors exactly corresponding to the list of five different models reported in Table \ref{tab:parameters}. 
Note that the change of slope, for all spectra, obviously correspond to the frequency $f_s$ reported in Eq. (\ref{11}) and obtained with the fit of the NANOGrav data (the corresponding  spectral amplitude is different, of course, as it describes primordial scalar perturbations).

It should be stressed, finally, that such a class of scalar spectra is well compatible with present phenomenological bounds (see e.g. \cite{Green}, \cite{Yi}), and may have interesting applications on the possible production of primordial black holes (PBH), as we are planning to discuss in a future paper.

\section{Conclusion} 
\label{sec4}

 The conclusion of this  paper is that the interpretation of the PTA signal \cite{2,2a,2b,2c,2d,2e,2f} as due to a stochastic background of relic gravitons produced in a string model of pre-big bang evolution, as suggested in \cite{1} on the ground of a  simple but phenomenologically incomplete GW spectrum, and referred in particular to a fit of the NANOGrav 15-year data set, may still be valid, and consistent with all constraints, if based on GW spectra obtained from more general, ``non-minimal" string cosmology scenarios. 
 
 In this paper we have followed the results of \cite{1}, and we assumed ab initio that the frequency obtained with the fit of the  NANOGrav data can be identified with the frequency $f_s$ of the transition from the dilaton to the string-curvature phase (see Eq. (\ref{11})). One could ask what happens if $f_s$ is left as another parameter of the model to be best-fitted to the data. While we are convinced that choosing $f_s \gg f_{\text{PTA}} \approx 1.2 \times 10^{-8} \text{Hz} $ would not be able to reproduce observations (because of the fast drop of the spectrum below $f_s$) it is not clear to what extent the case $f_s < f_{\text{PTA}}$ can be ruled out. 
We could ask, also, to what extent the results of this paper may change if, instead of starting from the fit of Eq. (\ref{11}), one would attempt a global fit of the complete set of PTA data.
 Answering these questions would amount to finding out to what extent our non-minimal model is fine-tuned.
 
 We are planning to present in a forthcoming paper a more detailed discussion of the possible values of the parameters of the non-minimal scenario and of their physical implications. In particular, we will give a possible string-theoretical  interpretation of the approximately flat  GW spectra which seems to be needed just above $f_s$, if we want to attribute the PTA signal to a cosmological scenario of this type.

\section*{Note Added} 
While in the final stage of writing this paper we noticed a new preprint \cite{Tan:2024urn} by the authors of Ref. \cite{1} in which their previous claims are reconsidered precisely in the minimal model of our Sect. 2 and ref. \cite{7}.
Their conclusion, that the minimal model cannot explain their fit of the NANOGrav data,  agrees with both \cite{7} and this paper. However, their claim that one can fit the data by relaxing slightly the upper bound in (\ref{25}) is incorrect. We thank the authors of \cite{Tan:2024urn} for confirming this in a private communication and for promptly posting a revised version.
\section*{Acknowledgements} 

We are grateful to Gianluca Calcagni for useful comments and for providing us with the sensitivity data of the gravitational antennas illustrated in Fig.~\ref{f2a}. 
GF acknowledges support by FCT under the research project number  PTDC/FIS-AST/ 0054/2021. GF is also member of the Gruppo Nazionale per la Fisica Matematica (GNFM) of the Istituto Nazionale di Alta Matematica (INdAM).
MG, EP and LT are supported in part by INFN under the program TAsP: {\it ``Theoretical Astroparticle Physics"}. EP and LT are also supported by the research grant number 2022E2J4RK ``PANTHEON: Perspectives in Astroparticle and Neutrino THEory with Old and New messengers", under the program PRIN 2022 funded by the Italian Ministero dell'Universit\`a e della Ricerca (MUR) and by the European Union-Next Generation EU. EP and LT want to acknowledge the hospitality of CERN, Department of Theoretical Physics. PC is supported in part by INFN under the program InDark: {\it ``Inflation, Dark Matter and the Large-Scale Structure of the Universe"}. PC also acknowledges support by the program PRIN 2022
- grant 20228RMX4A, funded by MUR and by the European Union - Next generation EU, Mission 4, Component 1, CUP C53D23000940006.
%

\appendix
\section{Useful relations among the parameters of the non-minimal model}
\label{secA}

In this appendix 
we provide some useful explicit relations that can be found by solving the set of equations (\ref{35}), (\ref{39}), (\ref{310}), (\ref{311}).

The first set of relations (\ref{35}), (\ref{39}) gives the five quantities $ H_1, m, z_s, z_d, z_\sg $ in terms of ${\sigma_i}/{\Mp}$, $\beta_h$ and $\beta_{\sg}$:

\bea
 && \log z_s = - \frac{K}{(\beta_{\sg}-\beta_h)}~; ~~~~~~~~~~~~~~~~~~~~~~~~~~~~~~~~~~~~~~~~~~~~~~~~~\log \frac{H_1}{\Mp} = \frac{C}{2} - \frac{K \beta_h}{(\beta_{\sg}-\beta_h)};\nonumber \\  &&
 \log \frac{z_s}{z_{\sg}}  = \frac32 B - \frac58 C + 3 \log\left(\frac{\sigma_i}{\Mp}\right) - \frac54 \frac{K(2- \beta_h)}{(\beta_{\sg}-\beta_h)}; ~~~~~~~~~~~~ 
 \log \frac{z_d}{z_{\sg}} = -2 B + \frac12 C + \frac{K(2- \beta_h)}{(\beta_{\sg}-\beta_h)}~; \nonumber \\
 && \log \frac{m}{\Mp} = 3B - \frac34 C + 2 \log\left(\frac{\sigma_i}{\Mp}\right) - \frac32 \frac{K(2- \beta_h)}{(\beta_{\sg}-\beta_h)}~,
 \label{rel1}
\eea
where we have defined the following (in general ${\sigma_i}/{\Mp}$-dependent) quantities:
\begin{align}
 & K\left(\frac{\sigma_i}{\Mp}\right) = \frac12\left[ A\left(\frac{\sigma_i}{\Mp}\right) - C +(n_s-1) B \right] ~,\nonumber \\
 &A = \log \left(\frac{4.2\pi^2}{T^2(\sg_i)} \right) -9 +(1-n_s) (\log 1.5 - 27) ~,\nonumber\\
 &B=\log \left( \frac{2\pi f_s}{H_0^{\frac{1}{2}} \Mp^{\frac{1}{2}}} \right)  - \frac{1}{6} \log \left( \frac{H_0}{H_{\mathrm{eq}}} \right) =\log\left(\frac{2\pi f_s}{3.9\times 10^{11}}\right) ~, \nonumber \\
 & C = \log \frac{\Omega_{\textsc{gw}}(f_s)} {\Omega_r(t_0)} = -4 + \log 2.9~.
 \label{def1}
\end{align}

Eqs. (\ref{rel1}) show that  the two ratios ${z_s}/{z_\sigma} $, 
${z_s}/{z_d} $ (and therefore also ${z_d}/{z_\sigma} $) can be expressed in terms of ${\sigma_i}/{\Mp}$, $m/\Mp, $ and other known constants. More explicitly we find:
\begin{align}
\log \left( \frac{z_s}{z_d} \right) &= \frac{3}{2} \log \left( \frac{m}{\Mp} \right) + \frac{1}{6} \log \left( \frac{H_0}{H_{\mathrm{eq}}} \right) + \log \left( \frac{H_0^{\frac{1}{2}} \Mp^{\frac{1}{2}}}{2\pi f_s} \right) \nonumber\\
&= \frac{3}{2} \log \left( \frac{m}{\Mp} \right) + \log \left( \frac{3.9 \times 10^{11}}{2\pi f_s} \right)\,,\\
\nonumber\\
\log \left( \frac{z_s}{z_\sigma} \right) &= \,
\frac{5}{6} \log \left( \frac{m}{\Mp} \right) + \frac{4}{3} \log \left( \frac{\sigma_i}{\Mp} \right) + \frac{1}{6} \log \left( \frac{H_0}{H_{\mathrm{eq}}} \right) + \log \left( \frac{H_0^{\frac{1}{2}} \Mp^{\frac{1}{2}}}{2\pi f_s} \right) \nonumber\\
&= \,
\frac{5}{6} \log \left( \frac{m}{\Mp} \right) + \frac{4}{3} \log \left( \frac{\sigma_i}{\Mp} \right) + \log \left( \frac{3.9 \times 10^{11}}{2\pi f_s} \right)\,,\\
\nonumber\\
\log \left( \frac{z_d}{z_\sigma} \right) &= \frac{4}{3} \log \left( \frac{\sigma_i}{\Mp} \right) - \frac{2}{3} \log \left( \frac{m}{\Mp} \right),
\end{align}
where $H_\text{eq}=1.6 \times 10^5 H_0=9.5 \times 10^{-56}\Mp$ and $\Mp \approx 2 \times 10^{18}\, \text{GeV}$.


\begin{thebibliography}{999}
\newcommand{\bb}{\bibitem}




\bb{2}\au{G}{Agazie} {et al.} [NANOGrav], \tia{The NANOGrav 15-year data set: evidence for a gravitational-wave background} \doinn{10.3847/2041-8213/acdac6}{Astrophys.\ J.\ Lett.}{951}{L8}{2023} [\arX{2306.16213}].

\bb{2a}\au{G}{Agazie} {et al.} [NANOGrav], \tia{The NANOGrav 15 yr Data Set: Observations and
Timing of 68 Millisecond Pulsars} \doinn{10.3847/2041-8213/acda9a}{Astrophys. J. Lett.}{951} {2023} {L9} [\arX{2306.16217}].

\bb{2b}\au{A} {Zic} {et al.}, \tia{The Parkes Pulsar Timing Array third data release} 
\doinn{10.1017/pasa.2023.36 }{Publ. Astron. Soc.
Austral.} {40} {2023} {e049} [\arX{2306.16230}].

\bb{2c}\au{DJ} {Reardon} {et al.}, \tia{Search for an Isotropic Gravitational-wave Background with the
Parkes Pulsar Timing Array}\doinn{10.3847/2041-8213/acdd02} {Astrophys. J. Lett.} {951} {2023} {L6} [\arX{2306.16215}].

\bb{2d}\au{J}{Antoniadis} [EPTA], \tia{The second data release from the European Pulsar Timing Array
- I. The dataset and timing analysis}, 
\doinn{10.1051/0004-6361/202346841 }
{Astron. Astrophys.} {678} {2023} {A48} [\arX{2306.16224}].

\bb{2e}[EPTA, InPTA], \tia{The second data release from the European Pulsar
Timing Array - III. Search for gravitational wave signals} \doinn{10.1051/0004-6361/202346844 } {Astron. Astrophys.} {678} 
{2023} {A50} [\arX{2306.16214}].

\bb{2f} \au{H} {Xu} {et al.}, \tia{Searching for the Nano-Hertz Stochastic Gravitational Wave Background
with the Chinese Pulsar Timing Array Data Release I}, 
\doinn{10.1088/1674-4527/acdfa5}{Res. Astron. Astrophys.} {23}
{2023} {075024} [\arX{2306.16216}].

\bb{1}Q. Tan, Y. Wu and L. Liu, \tia{Constraining string cosmology with the gravitational-wave background using the NANOGrav 15-year data set} [\arX{2409.17846}]. 



\bb{3} \au{M}{Gasperini} and \au{G}{Veneziano}, \tia{Pre-big bang in string cosmology} \doinn{10.1016/0927-6505(93)90017-8}{Astropart.\ Phys.}{1}{317}{1993} [\oarX{hep-th/9211021}].



\bb{4} \au{M}{Gasperini} and \au{G}{Veneziano}, \tia{The pre-big bang scenario in string cosmology} \doinn{10.1016/S0370-1573(02)00389-7}{Phys.\ Rept.}{373}{1}{2003} [\oarX{hep-th/0207130}].

\bibitem{Lidsey:1999mc}
J.~E.~Lidsey, D.~Wands and E.~J.~Copeland,
\tia{Superstring cosmology}\doinn{10.1016/S0370-1573(00)00064-8}
{Phys. Rept.} {337}{343} {2000} 
 [\oarX{hep-th/9909061}].

\bibitem{5} \au{M}{Gasperini} and \au{G}{Veneziano}, \tia{String theory and pre-big bang cosmology} \doin{10.1393/ncc/i2015-15160-8}{Nuovo Cim.}{C}{38}{160}{2016} [\oarX{hep-th/0703055}]. 




\bb{6} \au{G}{Veneziano},
\tia{Scale factor duality for classical and quantum strings} \doin{10.1016/0370-2693(91)90055-U}{Phys.\ Lett.}{B}{265}{287}{1991}. 


\bibitem{Meissner:1991zj}
K.~A.~Meissner and G.~Veneziano,
\tia{Symmetries of cosmological superstring vacua}
\doin{10.1016/0370-2693(91)90520-Z}
{Phys. Lett.} {B} {267} {33} {1991}

\bibitem{Meissner:1991ge}
K.~A.~Meissner and G.~Veneziano,
\tia{Manifestly O(d,d) invariant approach to space-time dependent string vacua} \doin{10.1142/S0217732391003924}
{Mod. Phys. Lett.} {A} {6}{3397} ({991}
[\oarX{hep-th/9110004}].
%

\bibitem{Gasperini:1991ak}
M.~Gasperini and G.~Veneziano,
\tia{O(d,d) covariant string cosmology} 
\doin{10.1016/0370-2693(92)90744-O}
{Phys. Lett.} {B} {277}{256} {1992}
[\oarX{hep-th/9112044}].


\bibitem{Sen:1991zi}
A.~Sen,
\tia{O(d) x O(d) symmetry of the space of cosmological solutions in string theory, scale factor duality and two-dimensional black holes}
\doin{10.1016/0370-2693(91)90090-D}
{Phys. Lett.} {B} {271}{295} {1991}. 

\bb{12}E~J.~Copeland, J.~E.~Lidsey and D.~Wands,
\tia{S-duality invariant perturbations in string cosmology} \doin{10.1016/S0550-3213(97)00538-5}{Nuc. Phys.} {B}{506}{407}{1997}
[\oarX{hep-th/9705050}].

\bb{7}
I.~Ben-Dayan, G.~Calcagni, M.~Gasperini, A.~Mazumdar, E.~Pavone, U.~Thattarampilly and A.~Verma,
\tia{Gravitational-wave background in bouncing models from semi-classical, quantum and string gravity} 
 \doij{10.1088/1475-7516/2024/09/058} {JCAP} {09} {058} {2024} 
  [\arX{2406.13521}]. 

\bb{8} \au{M}{Gasperini} and \au{M}{Giovannini}, \tia{Dilaton contributions to the cosmic gravitational wave
background} \doin{10.1103/PhysRevD.47.1519}{Phys.\ Rev.}{D}{47}{1519} {1993} [\oarX{gr-qc/9211021}]. 

\bb{9} \au{R}{Brustein}, \au{M}{Gasperini}, \au{M}{Giovannini} and \au{G}{Veneziano}, \tia{Relic gravitational waves from
string cosmology} \doin{10.1016/0370-2693(95)01128-D}{Phys.\ Lett.}{B}{361}{45}{1995} [\oarX{hep-th/9507017}].

\bb{10} \au{R}{Brustein}, \au{M}{Gasperini} and \au{G}{Veneziano}, \tia{Peak and endpoint of the relic graviton background in string cosmology} \doin{10.1103/PhysRevD.55.3882}{Phys.\ Rev.}{D}{55}{3882}{1997} [\oarX{hep-th/9604084}]. 

\bb{11}M.~Gasperini,
\tia{Elementary introduction to pre-big bang cosmology
 and to the relic graviton background} in \tia{Gravitational Waves,  Proc. of the Second SIGRAV School  on
 ``Gravitational Waves in  Astrophysics, Cosmology and String Theory"} 
(Centre A. Volta, Como, April 1999),
eds. I. Ciufolini et al. (IOP Publishing, Bristol, 2001), p. 280.
[\oarX{hep-th/9907067}].


\bb{13}V.~Bozza, M.~Gasperini, M.~Giovannini and G.~Veneziano,
\tia{Assisting pre-big bang phenomenology through short lived axions}
\doin{10.1016/S0370-2693(02)02387-0}
{Phys. Lett.} {B}{543}{14} {2002}
[\oarX{hep-ph/0206131}].

\bb{14}\au{V}{Bozza}, \au{M}{Gasperini}, \au{M}{Giovannini} and \au{G}{Veneziano}, \tia{Constraints on pre-big bang parameter space from CMBR anisotropies} \doin{10.1103/PhysRevD.67.063514}{Phys.\ Rev.}{D}{67}{063514}{2003} [\oarX{hep-ph/0212112}]. 

\bb{15} \au{M}{Gasperini}, \tia{Elements of String Cosmology} {Cambridge University Press} {Cambridge} {UK} {(2007)}.

\bb{16}K. Enqvist and M.S. Sloth, \tia{Adiabatic CMB perturbations in pre-big bang string cosmology} \doin{10.1016/S0550-3213(02)00043-3}
{Nucl. Phys.} {B} {626} {395} {2002}  [\oarX{hep-ph/0109214}].


\bb{17}D. H. Lyth and D. Wands, \tia{Generating the curvature perturbation without an inflaton} \doin{10.1016/S0370-2693(01)01366-1}
{Phys. Lett.} {B} {524}{5}{2002}  [\oarX{hep-ph/0110002}].

\bb{18} \au{M}{Gasperini}, \tia{Observable gravitational waves in pre-big bang cosmology: an update} \doij{10.1088/1475-7516/2016/12/010}{JCAP}{1612}{010}{2016} [\arX{1606.07889}]. 

\bibitem{19}P. Conzinu, M. Gasperini and G. Marozzi,
\tia{Primordial black holes from
pre-big bang inflation} 
 \doij{doi.org/10.1088/1475-7516/2020/08/031}
{JCAP} {08}{2020} {031}    [\arX{2004.08111}]. 

\bibitem{Conzinu:2023fth}\au{P}{Conzinu}, \au{G}{Fanizza}, \au{M}{Gasperini}, \au{E}{Pavone}, \au{L}{Tedesco} and \au{G}{Veneziano}, \tia{From the string vacuum to FLRW or de Sitter via $\alpha'$ corrections} \doij{10.1088/1475-7516/2023/12/019}{JCAP}{2312}{019}{2023} [\oarX{2308.16076}]. 

\bibitem{Conzinu:2023fui}
P.~Conzinu and G.~Marozzi,
\tia{Primordial black holes formation in an early matter dominated era from the pre-big-bang scenario} 
\doin{10.1103/PhysRevD.108.043533}
{Phys. Rev.} {D} {108} {2023} {043533}
[\oarX{2305.01430}].

\bb{29a}M. Gasperini and G. Veneziano, \tia{Dilaton production in string cosmology} \doin{10.48550/arXiv.gr-qc/9403031}{Phys. \ Rev.}{D}{50}
{1994}{2519} [\oarX{gr-qc/9403031}].

\bb{29b}R. Brustein, M. Gasperini, M. Giovannini, V.F. Mukhanov and G. Veneziano, \tia{Metric
perturbations in dilaton driven inflation} \doin{10.1103/PhysRevD.51.6744} 
{Phys. \ Rev.}{D}{51}{1995}{6744} [\oarX{hep-th/9501066}]. 

\bb{19a}V.F. Mukhanov, H.A. Feldman and R.H. Brandenberger \tia{Theory of cosmological perturbations}\doij{10.1016/0370-1573(92)90044-Z}
{Phys. Rept.} {215} {203}{1992}.

\bb{20}M. Gasperini, M. Maggiore and G. Veneziano, \tia{Towards a nonsingular pre-big bang cosmology} \doin{10.1016/S0550-3213(97)00149-1}
{Nucl. Phys.} {B} {494}{315} {1997}  [\oarX{hep-th/9611039}].

\bb{21} \au{S}{Kawai}, \au{M.-a}{Sakagami} and \au{J}{Soda}, \tia{Instability of one loop superstring cosmology} \doin{10.1016/S0370-2693(98)00925-3}{Phys.\ Lett.}{B}{437}{284}{1998} [\oarX{gr-qc/9802033}]. 

\bb{32a}PARTICLE DATA GROUP collaboration, \tia{Review of Particle Physics} 
\doinn{10.1093/ptep/ptac097}
{PTEP}{2022}{2022}{083C01}


\bibitem{Veneziano:2003sz}
G.~Veneziano,
\tia{A Model for the big bounce}
\doinn{10.1088/1475-7516/2004/03/004}
{JCAP} {03} {2004} {004}
[\oarX{hep-th/0312182}].

\bb{22}J.~Quintin, R.~H.~Brandenberger, M.~Gasperini and G.~Veneziano,
\tia{Stringy black-hole gas in \ensuremath{\alpha}'-corrected dilaton gravity} 
 \doin{10.1103/PhysRevD.98.103519}
{Phys. Rev.} {D} {98}  {103519} {2018}
[\arX{1809.01658}]. 


\bibitem{Bitnaya:2023vda}
D.~Bitnaya, P.~Conzinu and G.~Marozzi,
\tia{On the stability of string-hole gas}
\doinn{10.1088/1475-7516/2024/01/025}
{JCAP} {01} {2024}{025}
[\oarX{2308.16764}].

\bb{Dolgov}G.~Fanizza, M.~Gasperini, E.~Pavone and L.~Tedesco,
\tia{Linearized propagation equations for metric fluctuations in a general 
(non-vacuum) background geometry}
\doinn{10.1088/1475-7516/2021/07/021}
{JCAP} {07} {2021}{021} [\arX{2105.13041}].

\bibitem{31} \au{R}{Abbott} {et al.} [LIGO Scientific, VIRGO and KAGRA], \tia{GWTC-3: compact binary coalescences observed by LIGO and Virgo during the second part of the third observing run} \doin{10.1103/PhysRevX.13.041039}{Phys.\ Rev.}{X}{13}{041039}{2023} [\arX{2111.03606}].

\bb{ska}G.~Janssen, G.~Hobbs, M.~McLaughlin, C.~Bassa, A.~Deller, M.~Kramer, K.~Lee, C.~Mingarelli, P.~Rosado, S.~Sanidas, A.~Sesana, L.~Shao, I.~Stairs, B.~Stappers, and J.~Verbiest,
\tia{Gravitational Wave Astronomy with the SKA}
\doin{10.22323/1.215.0037}
{Proc. of Sci.}{}{037}{}{2015}.

\bibitem{24a} \au{G}{Agazie} \textit{et al.} [International Pulsar Timing Array], \tia{Comparing recent PTA results on the nanohertz stochastic gravitational wave background} \arX{2309.00693}.


\bibitem{27} \au{P}{Auclair} {et al.} [LISA Cosmology Working Group], \tia{Cosmology with the Laser Interferometer Space Antenna} \doinn{10.1007/s41114-023-00045-2}{Living Rev.\ Relativ.}{26}{5}{2023} [\arX{2204.05434}].

\bibitem{28} \au{M}{Branchesi} {et al.} \tia{Science with the Einstein Telescope: a comparison of different designs} 
		\doij{10.1088/1475-7516/2023/07/068}{JCAP}{2307}{068}{2023} [\arX{2303.15923}].

\bibitem{29} \au{S}{Kawamura} {et al.}, \tia{Current status of space gravitational wave antenna DECIGO and B-DECIGO} \doinn{10.1093/ptep/ptab019}{Prog.\ Theor.\ Exp.\ Phys.}{2021}{05A105}{2021} [\arX{2006.13545}].


\bb{30}\au{R}{Abbott} {et al.} [LIGO Scientific, VIRGO and KAGRA], \tia{Tests of general relativity with GWTC-3} [\arX{2112.06861}].

\bb{Green}Ann B. Green, \tia{Primordial black holes as a dark matter candidate - a brief overview} \arX{2402.15211}

\bb{Yi}Zhu Yi and Qin Fei,
\tia{Constraints on primordial curvature spectrum from primordial black holes and scalar-induced gravitational waves}
\doinn{10.1140/epjc/s10052-023-11233-3}   {Eur. Phys. J}{C}{83}{2023} [\oarX{2210.0641}]. 


\bibitem{Tan:2024urn}
Q.~Tan, Y.~Wu and L.~Liu,
\tia{Pre-Big-Bang Cosmology Cannot Explain NANOGrav 15-year Signal}
[\oarX{2411.16505}].


 
%
\end{thebibliography}
\end{document}